\documentclass[12pt]{article}
\usepackage{epsfig}
\begin{document}
\begin{center}
{\Large {\bf $Z_6$ symmetry of the Standard Model and Technicolor theory}

\vskip-40mm \rightline{\small ITEP-LAT/2007-11} \vskip 30mm

{
\vspace{1cm}
{ M.A.~Zubkov$^{a,b}$ }\\
\vspace{.5cm} {\it $^a$ ITEP, B.Cheremushkinskaya 25, Moscow, 117259, Russia }
\\ \vspace{.5cm} {\it $^b$ Moscow Institute of Physics and Technology, 141700,
Dolgoprudnyi, Moscow Region, Russia  }}}

\end{center}

\begin{abstract}
We consider the possibility to continue the $Z_6$ symmetry of the Standard
Model to the Technicolor theories. Among the $SU(N)$ Weinberg - Susskind models
and the $SU(N)$ Farhi - Susskind models for $N>2$ only the $SU(4)$ Farhi -
Susskind model may possess the mentioned symmetry. We also show that the
hypercharge assignment of Minimal Walking $SU(2)$ Technicolor model may be
chosen in such a way that the additional discrete symmetry is preserved.
\end{abstract}

\section{Introduction}

The Standard Model with the fundamental scalar Higgs field is known to have
serious problems at the energies of the order of $1$ Tev. From the point of
view of perturbation theory this  scale appears in the so-called Hierarchy
problem \cite{TEV}. Namely, the mass $m^2$ for the scalar field receives the
quadratically divergent contribution in one loop. Therefore, formally the
initial mass parameter ($m^2= - \lambda v^2$, where $v$ is the vacuum average
of the scalar field while $\lambda$ is its self - coupling) should be set to
infinity in such a way that the renormalized mass $m^2_R$ remains negative and
finite. This is the content of the so-called fine tuning. It is commonly
believed that this fine tuning is not natural \cite{TEV} and, therefore, the
finite ultraviolet cutoff $\Lambda$ should be set up. From the requirement that
the one-loop contribution to $m^2$ is less than $10 |m_R^2|$ one derives that
 $\Lambda \sim 1$ TeV. This problem appears also in lattice nonperturbative
 study (see, for example, \cite{BVZ2007,VZ2008}). Thus it is natural to construct the new theory,
which describes Tev - scale physics and provides the spontaneous breakdown of
Electroweak symmetry.

QCD is usually considered as the self consistent quantum field theory contrary
to the Weinberg - Salam model. Therefore it is natural to construct the new Tev
scale theory basing on the analogy with QCD. This program is realized in the so
- called Technicolor theory\cite{technicolor, WS,FS}. Namely, the new
Nonabelian gauge interaction is added with the scale $\Lambda_{TC} \sim 1$ Tev,
where $\Lambda_{TC}$ is the analogue of $\Lambda_{QCD}$. This new interaction
is called Technicolor. The correspondent new fermions are called
technifermions. The Electroweak gauge group acts on the technifermions.
Therefore, breaking of the chiral symmetry in Technicolor theory causes
Electroweak symmetry breaking. This makes three of the four Electroweak gauge
bosons massive. However, pure Technicolor theory cannot explain appearance of
fermion masses.

In order to make Standard Model fermions massive extra gauge interaction may be
added, which is called Extended Technicolor (ETC)
\cite{technicolor,ExtendedTechnicolor}. In this new gauge theory the Standard
Model fermions and technifermions enter the same representation of the Extended
Technicolor group. Unfortunately, the first ETC models suffer from the
extremely large flavor - changing amplitudes and unphysically large
contribution to the Electroweak polarization operators \cite{technicolor}. The
way to overcome these problems is related to the behavior of chiral gauge
theories at large number of fermions or for the high order representations.
Namely, the near conformal behavior of the Technicolor model allows to suppress
dangerous flavor changing currents as well as to decrease the contribution to
the S - parameter \cite{Appelquist,minimal_walking}. (It is worth mentioning,
however, that the generation of $t$ - quark mass in these models still causes
serious problems.)

There is a great number of Technicolor and Extended Technicolor models. That's
why it is important to find a general principle, which may help to make a
choice. The present paper is an attempt to extract such a principle from the
additional $Z_6$ symmetry of the Standard Model. It has been found long time
ago, that the spontaneous breakdown of $SU(5)$ symmetry in Grand Unified Theory
actually leads to the gauge group $SU(3)\times SU(2) \times U(1)/Z_6$ instead
of the conventional $SU(3)\times SU(2) \times U(1)$ (see, for example,
\cite{Z6} and references therein). However, the $Z_6$ symmetry is not the
subject of the $SU(5)$ unification only. Actually, the $Z_6$ symmetry is
present in the Standard Model itself without any relation to the particular
Unified theory \cite{BVZ2003,BVZ2004, BVZ2005, BVZ2006,Z6f}. The $Z_6$ symmetry
is  rather restrictive and it forbids, for example, the appearance of such
particles as left - handed Standard Model fermions with zero hypercharge. It
was shown in \cite{BVZ2003}, that the Unified models based on the Pati - Salam
scheme may possess the $Z_6$ symmetry. Besides, it was found that in the so -
called Petite Unification models (also based on the Pati-Salam scheme) the
additional discrete symmetry is present ($Z_2$ or $Z_3$ depending on the choice
of the model)  \cite{Z2007}.

Here we suggest the way to continue the mentioned $Z_6$ symmetry to Technicolor
theories. Then we require that the Technicolor models possess this additional
discrete symmetry and find, that this requirement gives an essential
restriction on the choice of Technicolor theory. As it was shown in
\cite{Z2007}, the observability of the $Z_6$ symmetry of the Standard Model is
related to the monopole content of the Unified theory. We do not need to know
the details of the Unified model. The appearance of the additional $Z_6$
symmetry restricts essentially the monopole content of the Unified theory. In
this paper we demonstrate that the same is valid for the theory that unifies
Technicolor theory with the Standard Model. Usually the role of the Unified
theory for Technicolor and Standard Model interactions is played by Extended
Technicolor. Thus the additional discrete symmetry of the Technicolor should
restrict the monopole content of Extended Technicolor.

The paper is organized as follows.

In  section $2$ we remind the reader the content of the additional $Z_6$
symmetry of the Standard Model.

In  section $3$ we suggest the way to continue the $Z_6$ symmetry of the
Standard Model to its Technicolor extension.

In  section $4$ we consider the possibility to continue the given additional
discrete symmetry to the minimal Technicolor model by Weinberg and Susskind.

In section $5$ we consider in the same context Technicolor model by Farhi and
Susskind.

In section $6$ we consider the relation between the additional discrete
symmetry of the Technicolor models and the monopole content of the Extended
Technicolor models.

In section $7$ we show that minimal walking Technicolor model may be
constructed in such a way, that it preserves the additional $Z_6$ symmetry.

In section $8$ we end with the conclusions.

\section{$Z_6$ symmetry}

Here we remind the reader of what we call the additional $Z_6$ symmetry in the
Standard Model.

For any path $\cal C$,  we may calculate the elementary parallel transporters
\begin{eqnarray}
\Gamma &=& {\rm P} \, {\rm exp} (i\int_{\cal C} C^{\mu} dx^{\mu}) \nonumber\\
U &=& {\rm P} \, {\rm exp} (i\int_{\cal C} A^{\mu} dx^{\mu}) \nonumber\\
e^{i\theta} &=& {\rm exp} (i\int_{\cal C} B^{\mu} dx^{\mu}) ,\label{Sing}
\end{eqnarray}
where $C$, $A$, and $B$ are correspondingly $SU(3)$, $SU(2)$ and $U(1)$ gauge
fields of the Standard Model.

The parallel transporter correspondent to each fermion of the Standard Model
(or to the scalar Higgs) is the product of the elementary ones listed above.
Therefore,  the elementary parallel transporters are encountered in the theory
only in the combinations listed in the table.

\begin{table}
\label{tab.01}
\begin{center}
\begin{tabular}{|c|l|}
\hline
$U\, e^{-i\theta}$ & {\rm left-handed leptons} \\
\hline
$e^{-2 i \theta}$ & {\rm right-handed leptons} \\
\hline
$ \Gamma \, U \, e^{ \frac{i}{3} \theta}$ & {\rm left-handed quarks}\\
\hline
$ \Gamma \, e^{ -\frac{2i}{3} \theta}$ &{\rm right-handed $d$, $s$, and, $b$ - quarks} \\
\hline
$ \Gamma \, e^{ \frac{4i}{3} \theta}$ &{\rm right-handed $u$, $c$, and, $t$ - quarks} \\
\hline
$  U \, e^{  i \theta}$ &{\rm the Higgs scalar field}\\
\hline
\end{tabular}
\end{center}
\end{table}
It can be easily seen \cite{BVZ2003} that {\it all} the listed combinations are
invariant under the following $Z_6$ transformations:
\begin{eqnarray}
 U & \rightarrow & U e^{i\pi N}, \nonumber\\
 \theta & \rightarrow & \theta +  \pi N, \nonumber\\
 \Gamma & \rightarrow & \Gamma e^{(2\pi i/3)N},
\label{symlat}
\end{eqnarray}
where $N$ is an arbitrary integer number.  This symmetry allows to define the
Standard Model with the gauge group $SU(3)\times SU(2) \times U(1)/{\cal Z}$
(${\cal Z} = Z_6$, $Z_3$ or $Z_2$) instead of the usual $SU(3)\times SU(2)
\times U(1)$.

As it was mentioned in the introduction, the fact that the Standard Model may
have the gauge group $SU(3)\times SU(2) \times U(1)/Z_6$ can be recognized
during the consideration of the $SU(5)$ unified model \cite{Z6}. The $SU(5)$
parallel transporter at low energies has the form
\begin{equation}
\Omega = \left( \begin{array}{c c}

\Gamma^+ e^{\frac{2i\theta}{3}} & 0  \\
0 & Ue^{-i\theta}

\end{array}\right)\in SU(5), \label{SU(5)}
\end{equation}
where $\Gamma, U$, and $e^{i\theta}$ are the elementary parallel transporters
of the Standard Model. (\ref{SU(5)}) is obviously invariant under
(\ref{symlat}), which means that the breakdown pattern is $SU(5) \rightarrow
SU(3)\times SU(2) \times U(1)/Z_6$, and not $SU(5) \rightarrow SU(3)\times
SU(2) \times U(1)$.

On the level of perturbation expansion around flat vacuum the mentioned
versions of the Standard Model are indistinguishable. In \cite{Z2007} it was
shown that the situation is changed if one considers the Standard Model in
space of nontrivial topology, or, to be embedded into the unified model with
the simply connected gauge group. Then the monopole content of the unified
theory is completely different for the mentioned versions of the Standard
Model. Petite Unification Theory \cite{PUT} gives an example of realistic
theory, in which the unification of interactions occurs at the Tev scale. In
this theory due to the additional discrete symmetry of the Standard Model the
topologically stable monopoles may appear with the masses of the order of $10$
Tev \cite{Z2007}.

\section{How to continue $Z_6$ symmetry to the Technicolor models}

It is worth mentioning that the additional discrete symmetry is rather
restrictive. Namely, for the Standard Model the requirement that the fermion
parallel transporters are invariant under $Z_6$ gives the condition for the
choice of the representations that are allowed for the Standard Model fermions.
Say, the left - handed $SU(2)$ doublets with zero hypercharge are forbidden.

The nature of the given additional  symmetry is related to the centers $Z_3$
and $Z_2$ of $SU(3)$ and $SU(2)$. This symmetry connects the centers of $SU(2)$
and $SU(3)$ subgroups of the gauge group. We suggest the following way to
continue this symmetry to the Technicolor extension of the Standard Model.

We connect the center of the  Technicolor group to the centers of $SU(3)$ and
$SU(2)$. Let $SU(N_{TC})$ be the Technicolor group. Then the transformation
(\ref{symlat}) is generalized to
\begin{eqnarray}
 U & \rightarrow & U e^{i\pi N}, \nonumber\\
 \theta & \rightarrow & \theta +  \pi N, \nonumber\\
 \Gamma & \rightarrow & \Gamma e^{(2\pi i/3)N},\nonumber\\
 \Theta & \rightarrow & \Theta e^{(2\pi i/N_{TC})N}.
\label{symlatWS}
\end{eqnarray}
Here $\Theta$ is the $SU(N_{TC})$ parallel transporter. The parallel
transporters correspondent to the new fermions of the theory should be
invariant under (\ref{symlatWS}). It should be mentioned that the resulting
symmetry is not the product of $Z_6$ and $Z_{N_{TC}}$.


\section{Minimal Technicolor model of Weinberg and Susskind} We consider here
the simplest Technicolor model by Weinberg and Susskind \cite{WS} (see also
\cite{technicolor} and references therein). The model contains technifermions
\begin{equation}
 \left(
 \begin{array}{c}
  T^a \\
  B^a
 \end{array}
 \right)_L , \quad
 \left(
 \begin{array}{c}
 T^a \\
  B^a
 \end{array}
 \right)_R .
\end{equation}
The hypercharge assignment is $Y=0$ for the left - handed technifermions and $Y
= \pm 1$ for the right - handed ones. Index $a$ corresponds to the Technicolor
group $SU(N_{TC})$. The model has local $SU(2)_L$ gauge symmetry and global
$SU(2)_R$ symmetry. Chiral symmetry breaking provides breakdown of Electroweak
symmetry and formation of massive $W$ and $Z$ bosons. One can also consider
$N_D \ne 1$ copies of technifermions.

Now we require that the parallel transporters correspondent to the
technifermions are invariant under (\ref{symlatWS}). This leads to the
following condition:

\begin{equation}
\frac{2\pi N}{N_{TC}} + \pi N = 2 \pi k(N), \, k(N)\in Z
\end{equation}

The only solution of this equation is $N_{TC} = 2, k(N) = N$. Thus we conclude
that the Technicolor model of Weinberg and Susskind is invariant under the
extention of $Z_6$ symmetry of the Standard Model if the Technicolor group is
$SU(2)$. The correspondent additional symmetry is $Z_{6}$. The given
Technicolor Extention of the Standard model is, therefore, allowed to have the
gauge group $SU(2)\otimes SU(3)\times SU(2) \times U(1)/Z_{6}$ instead of
$SU(2)\otimes SU(3)\times SU(2) \times U(1)$.

It is worth mentioning that the $SU(2)$ Technicolor interactions alone suffer
from the vacuum alignment problems. In order to demonstrate this let us remind
briefly the consideration of \cite{Align}. We define the field
\begin{equation}
 Q^{\dot{\alpha},a} = \left(
 \begin{array}{c}
  T_L^{\dot{\alpha},a} \\
  B_L^{\dot{\alpha},a}\\
 \epsilon^{a a^\prime} [T_R^{\alpha,a^\prime}]^* \\
  \epsilon^{a a^\prime} [B_R^{\alpha,a^\prime}]^* \\
   \end{array}
 \right)
\end{equation}
(Here $\alpha, \beta, \dot{\alpha}, \dot{\beta}$ are spinor indices. The
conventional four component spinor $T$ is  composed of the two component
spinors as $T = (T_R^{1}, T_R^{2}, \epsilon_{1 \alpha} T_L^{\dot{\alpha}},
\epsilon_{2 \alpha} T_L^{\dot{\alpha}})^T$.)

 $Q^{\dot{\alpha},a}$ transforms as a left - handed
dotted spinor under $SL(2,C)$ and as an element of fundamental representation
under Technicolor $SU(2)$. If the Electroweak interactions are switched off the
Technicolor lagrangian is invariant under the global $SU(4)$ symmetry. The
$SU(2)$ and $SL(2,C)$ invariant bilinear combination of $Q$ is
\begin{equation}
\Phi_{AB} = \epsilon_{ab}\epsilon_{\alpha \beta} Q^{\dot{\alpha},a}_A Q^{
\dot{\beta},b}_B,
\end{equation}
where $A,B$ are $SU(4)$ indices.

The low energy effective potential $V(\Phi)$ is invariant under the action of
$SU(4)$ on $\Phi$. The correct vacuum value of $\Phi$ is chosen when the
$SU(4)$ breaking perturbations are taken into account. This process is known as
vacuum alignment \cite{Align}.

The conventional Electroweak vacuum corresponds to the value of $\Phi$
proportional to
\begin{equation}
 \Phi  = \left(
 \begin{array}{c c c c}
  0 & 0 & 1 & 0 \\
  0 & 0 & 0 & 1 \\
 -1 & 0 & 0 & 0  \\
  0 & -1 & 0 & 0  \\
   \end{array}
 \right)
\end{equation}
The correspondent chiral condensate is $\langle \delta_{ab} \epsilon_{\alpha
\beta} [T^{\beta b}_R]^* T^{\dot{\alpha}a}_L  + \delta_{ab}\epsilon_{\alpha
\beta}[B^{ \beta b}_R]^* B^{\dot{\alpha} a}_L  \rangle$.

However, the following value is also admitted
\begin{equation}
 \Phi  = \left(
 \begin{array}{c c c c}
  0 & 1 & 0 & 0 \\
  -1 & 0 & 0 & 0 \\
 0 & 0 & 0 & 1  \\
  0 & 0 & -1 & 0  \\
   \end{array}
 \right)
\end{equation}
In this case the condensate is $\langle \epsilon_{ab}\epsilon_{\alpha \beta}
T^{\dot{\alpha}a}_L B^{ \dot{\beta}b}_L + \epsilon_{ab}\epsilon_{\alpha \beta}
[T^{\alpha a}_R]^+ [B^{ \beta b}_R]^+\rangle $ and the Electroweak $SU(2)$
remains unbroken while photon becomes massive.

In general the chiral symmetry breaking is a rather complicated phenomenon and
its physics is still not understood in sufficient details. Therefore, it is not
completely clear which one of the two mentioned possibilities is realized. The
analysis of \cite{Align} shows that the small perturbations due to the
Electroweak interactions choose such a vacuum that the Electroweak symmetry is
broken in a minimal way, i.e. the sum of squared gauge boson masses is
minimized. With this rule we come to the conclusion that in the $SU(2)$
Weinberg - Susskind model the second of the two possibilities mentioned above
is realized and the Electroweak symmetry is broken incorrectly.

\section{Farhi-Sasskind model}

The model \cite{FS,technicolor} contains four doublets
\begin{eqnarray}
 \left(
 \begin{array}{c}
  U_i^a \\
  D_i^a
 \end{array}
 \right)_L , \quad
 \left(
 \begin{array}{c}
 U_i^a \\
  D_i^a
 \end{array}
 \right)_R \nonumber\\
 \left(
 \begin{array}{c}
  N^a \\
  E^a
 \end{array}
 \right)_L , \quad
 \left(
 \begin{array}{c}
 N^a \\
  E^a
 \end{array}
 \right)_R .
\end{eqnarray}
Here $a$ is the Technicolor $SU(N_{TC})$ index while index $i$ corresponds to
the color group $SU(3)$. Colored fermions are called techniquarks, while the
others are called technileptons. The hypercharge of the left - handed
technileptons is denoted as $Y_L$. The hypercharges of the right - handed
technileptons are denoted as $Y^{1,2}_R$. The hypercharge of the left - handed
techniquarks is denoted as $Y^c_L$. The hypercharges of the right - handed
techniquarks are denoted as $(Y^c)^{1,2}_R$. The conventional chiral
condensates are invariant under the electromagnetic $U(1)$ transformations if
$Y^{1,2}_R = Y_L \pm 1$, $(Y^c)^{1,2}_R = Y^c_L \pm 1$.

The theory is anomaly - free if $Y_L + 3 Y^c_L = 0$. Let us now require that
the model is invariant under the additional symmetry (\ref{symlatWS}). Thus we
must have

\begin{eqnarray}
&&[\frac{2}{N_{TC}} + \frac{2}{3} + 1 + Y^c_L ] \, {\rm mod}\, 2 = 0 \nonumber\\
&&[\frac{2}{N_{TC}} +  1 + Y_L ] \, {\rm mod}\, 2 = 0\nonumber\\
&&   Y_L + 3 Y^c_L = 0
\end{eqnarray}

The given system of equations has the two sets of solutions:

1) $N_{TC} = 2, Y^c_L = -\frac{Y_L}{3}, Y_L = 2(1-3k), k \in Z$;

2) $N_{TC} = 4, Y^c_L = -\frac{Y_L}{3}, Y_L = \frac{1}{2} - 6k , k\in Z$.

Thus only groups $SU(2)$ and $SU(4)$ may serve as Technicolor groups of
Farhi-Sasskind model if we require that the theory possesses the additional
symmetry (\ref{symlatWS}). The following groups may be the gauge groups of the
correspondent extensions of the Standard Model:
\begin{equation}
SU(2)\otimes SU(3)\times SU(2) \times U(1)/Z_6
\end{equation}
or
\begin{equation}
SU(4)\otimes SU(3)\times SU(2) \times U(1)/Z_{12}
\end{equation}

It was already mentioned that the $SU(2)$ Technicolor interactions in the
Weinberg - Susskind model suffer from the vacuum alignment problems. In the
same way in the $SU(2)$ Farhi - Susskind model such problems appear in the
technilepton sector \cite{Align}. For this reason among the two models with
Technicolor groups $SU(2)$ and $SU(4)$ that preserve an additional discrete
symmetry the preferred Technicolor Farhi - Susskind model is the one with the
$SU(4)$ group.

\section{The unification of Technicolor and Standard Model interactions}

In this section we demonstrate how, in principle, the Technicolor and the
Standard Model interactions may be unified in a common gauge group.   We do not
discuss here the details of the breakdown mechanism and how the chiral anomaly
cancellation is provided within the given scheme of Unification. Our aim here
is to demonstrate how the additional discrete symmetry (\ref{symlatWS}) may
appear during the breakdown of Unified gauge symmetry.

For the definiteness let us consider $N_{TC} = 4$. Let $SU(10)$ be the Unified
gauge group. The breakdown pattern is $SU(10)\rightarrow SU(4)\otimes
SU(3)\times SU(2) \times U(1)/Z_{12}$. We suppose that at low energies the
$SU(10)$ parallel transporter has the form:

\begin{equation}
\Omega = \left( \begin{array}{c c c c}
\Theta e^{-\frac{2i\theta}{4}} & 0 & 0 & 0 \\
0 & \Gamma^+ e^{\frac{2i\theta}{3}} & 0 & 0 \\
0 & 0 & Ue^{-i\theta} & 0\\
0 & 0 & 0 & e^{2i\theta}
\end{array}\right)\in SU(10), \label{U(10)}
\end{equation}

The fermions of each generation $\Psi^{i_1 ... i_N}_{j_1 ... j_K}$  carry
indices $i_k$ of the fundamental representation of $SU(10)$ and the indices
$j_k$ of the conjugate representation. They may be identified with the Standard
Model fermions and Farhi - Susskind fermions as follows (we consider here the
first generation only):

\begin{eqnarray}
&& \Psi^{10} = e^c_R; \, \Psi_{10}^{10} = \nu_R; \, \Psi^{i_2} =
\left(\begin{array}{c} \nu_L
\\ e^-_L\end{array}\right) ;\nonumber\\
&&\Psi^{i_3} = d^c_{i_3,R} ; \, \Psi^{i_3}_{10} = u^c_{i_3,R} ;\, \Psi^{i_2}_{
i_3} =
\left(\begin{array}{c} u^{i_3}_L \\ d^{i_3}_L \end{array}\right) ;\nonumber\\
&&\Psi_{i_4} = E^c_{i_4,R} ; \, \Psi_{10, i_4} = N^c_{i_4,R} ;\, \Psi^{i_2 i_4}
=
\left(\begin{array}{c} N^{i_4}_L \\ E^{i_4}_L \end{array}\right) ;\nonumber\\
&&\Psi^{i_3}_{i_4} = D^c_{i_3 i_4,R} ; \, \Psi_{10, i_4}^{i_3} = U^c_{i_3
i_4,R} ;\, \Psi^{i_2 i_4}_{i_3} = \left(\begin{array}{c} U^{i_3 i_4}_L \\
D^{i_3 i_4}_L
\end{array}\right) \nonumber\\&& (i_2 = 8,9; \, i_3 =
5,6,7;\, i_4 = 1,2,3,4);
\end{eqnarray}

Here the charge conjugation is defined as follows: $f^c_{\dot{\alpha}} =
\epsilon_{\alpha \beta} [f^{\beta}]^*$.

In principle the fermion content of the Unified model should be chosen in such
a way that the anomalies are cancelled. Moreover, some physics should be added
in order to provide "unnecessary" fermions with the masses well above $1$ Tev
scale. Besides, one must construct the unambiguous theory in such a way that at
low energies the parallel transporters indeed have the form (\ref{U(10)}). All
these issues are to be the subject of an additional investigation. For now,
however, let us suppose that this program is fulfilled. Then all parallel
transporters in the theory are invariant under (\ref{symlatWS}) in a natural
way. The gauge group $SU(10)$ is simply connected. That's why the Unified
theory should contain monopole - like topological objects. The similar
situation was considered in \cite{Z2007}, where the Unification of the Standard
Model interactions was considered. In particular, monopole configurations with
the usual magnetic flux $2 \pi$ and hypercharge flux $\pi$ must be present.
(Electromagnetic field is expressed through the $SU(2)$ field $A$ and the
hypercharge field $B$ as follows: $A_{\rm em}  =  2 B + 2 \,{\rm sin}^2\,
\theta_W (A_3 - B)$.)
It is worth mentioning that the monopoles with the usual magnetic flux $ 4 \pi
\,{\rm cos}^2 \theta_W $ are also present in this theory.

Let us now suppose that the Unified theory is constructed in such a way that
the breakdown pattern is $G \rightarrow SU(4)\otimes SU(3)\times SU(2) \times
U(1)$. We also suppose that the hypercharges of the fermions are rational
numbers $\frac{P}{Q}$ with integer $P$ and $Q$, where the maximal value of $Q$
is $3$.  Then the monopole content of the Unified theory is essentially
different. Namely, the minimal hypercharge flux is $6 \pi$ while the usual
magnetic flux\footnote{If the maximal value of $Q$ is $Q_{max} > 3$ then the
magnetic flux of the monopole is  $ 4 Q_{max} \pi \,{\rm cos}^2 \theta_W $.} is
$ 12 \pi \,{\rm cos}^2 \theta_W $.


Thus, there is an essential difference between the monopole contents of the
Unified models in the two considered cases.  Namely, in the presence of
symmetry (\ref{symlatWS}), the monopoles with magnetic flux proportional to $2
\pi$ appear, while in the opposite case they do not appear.


The considered $SU(10)$ interactions may be included into the sequence of
Extended Technicolor interactions. Then the $SU(10)$ gauge group plays a role
in the fermion mass formation mechanism.

\section{Minimal walking Technicolor}

 Technicolor models with their chiral symmetry breaking
are able to provide breaking of Electroweak symmetry. But these models alone
cannot provide fermions with realistic masses.  Standard Model fermions become
massive if they may be transformed into technifermions, say, with ejecting of
the new massive gauge bosons. Then the quark and lepton masses are evaluated at
one loop level as $m_{q,l}\sim \frac{N_{TC}\Lambda_{TC}^3}{\Lambda_{ETC}^2}$,
where $\Lambda_{TC}$ is the Technicolor scale while $\Lambda_{ETC}$ is the
scale of the new strong interaction called Extended Technicolor. (Spontaneous
breakdown of Extended Technicolor symmetry gives rise to the mass of the new
gauge bosons
 of the order of $\Lambda_{ETC}$.)

The number of fermions (arranged in the fundamental representation of the
Technicolor group) for which the behavior of the model becomes conformal can be
evaluated \cite{Appelquist} as $N_f \sim 4 N_{TC}$. In this case the effective
charge becomes walking instead of running \cite{walking}. In the correspondent
ETC theory the flavor changing processes may be suppressed, which allows to
approach to the realistic description of the generation of the Standard Model
fermion masses. It is worth mentioning, however, that the realistic top quark
mass cannot be generated in this way without causing additional problems in the
theory. That's why the top quark mass generation should be a subject of an
additional efforts in ETC model - building (see, for example,
\cite{Sannino_t}). In the Farhi-Susskind model conformal regime is approached
for the number of technifermion generations $N_D$ equal to $N_D = 2$ at $N_{TC}
= 4$.

If the technifermions are arranged in the fundamental representation of the
walking Technicolor model, the perturbative contribution to S parameter still
remains dangerously large \cite{minimal_walking}. One way to avoid this problem
is to consider higher representations of Technicolor group. The minimal choice
here is $N_{TC} = 2$ with the one generation of technifermions from the two -
index symmetric representation of $SU(2)$. This minimal model contains
technifermions symmetric in Technicolor $SU(2)$ indices $a$ and $b$:
\begin{eqnarray}
 L^{a,b} = \left(
 \begin{array}{c}
  U^{a,b} \\
  D^{a,b}
 \end{array}
 \right)_L , \quad
 R^{a,b} =\left(
 \begin{array}{c}
 U^{a,b} \\
  D^{a,b}
 \end{array}
 \right)_R \nonumber\\
 \left(
 \begin{array}{c}
  N \\
  E
 \end{array}
 \right)_L , \quad
 \left(
 \begin{array}{c}
 N \\
  E
 \end{array}
 \right)_R .
\end{eqnarray}
Here extra generation of Standard Model leptons is added in order to cancel
chiral anomaly.  In this model the contribution to S - parameter is
sufficiently smaller than for the model with technifermions from the
fundamental representation \cite{minimal_walking}.

The anomaly is absent if $3 Y^c_L + Y_L = 0$, where $Y^c_L$ is the hypercharge
of the left-handed technileptons while $Y_L$ is the hypercharge of the new left
- handed leptons that are Technicolor singlets. It is important, that the given
two - index representation of Technicolor $SU(2)$ group does not feel the
center of $SU(2)$. Therefore, the parallel transporters correspondent to the
new fermions are invariant under (\ref{symlatWS}) with $N_{TC} = 2$ if
\begin{eqnarray}
 &&[1 + Y^c_L ]  {\rm mod} 2 = 0 \nonumber\\
 &&[1 + Y_L ]  {\rm mod} 2 = 0 \nonumber\\
&& Y_L + 3 Y^c_L = 0
\end{eqnarray}

The solution is $Y^c_L = -\frac{Y_L}{3}, Y_L = 3(1-2k), k \in Z$. Thus we
conclude, that the minimal walking Technicolor model can be made invariant
under the extension of the $Z_6$ symmetry of the Standard Model.

Let us also notice here that in  the given model the vacuum average $\langle
\epsilon_{cd}\epsilon_{ab}\epsilon_{\alpha \beta} U^{a,c,\dot{\alpha}}_L
D^{b,d, \dot{\beta}}_L + \epsilon_{cd}\epsilon_{ab}\epsilon_{\alpha \beta}
[U_R^{a,c,\alpha}]^+ [D_R^{b,d, \beta}]^+\rangle$ may appear instead of the
conventional  $\langle \epsilon_{cd}\epsilon_{ab}\epsilon_{\alpha \beta}
[U^{a,c,\dot{\alpha}}_L]^+ U^{b,d,
\beta}_R+\epsilon_{cd}\epsilon_{ab}\epsilon_{\alpha \beta}
[D^{a,c,\dot{\alpha}}_L]^+ D^{b,d, \beta}_R \rangle $. If so, the Electroweak
symmetry would not be broken properly. However, it can be shown \cite{Align},
that in this case the sum of squared gauge boson masses is larger, than for the
conventional breakdown. That's why the preferred vacuum orientation in this
case is the conventional one.

\section{Conclusions}

In this paper we suggest the way to continue the $Z_6$ symmetry of the Standard
Model to the Technicolor theory.

We have found that the minimal Technicolor model of Weinberg and Susskind may
possess the suggested additional discrete symmetry only for $N_{TC}=2$. In the
Farhi - Susskind model there are two possibilities: $N_{TC} = 2$ and
$N_{TC}=4$, for which the theory contains the additional discrete symmetry
(correspondingly, $Z_{6}$ and $Z_{12}$). In the latter case the complete theory
can be constructed with the gauge group $SU(4)\times SU(3)\times SU(2) \times
U(1)/Z_{12}$. It is worth mentioning that the $SU(2)$ Weinberg - Susskind and
Farhi - Susskind models suffer from the vacuum alignment problems. That's why
we do not consider them as realistic and the only possibility remains that is
the $SU(4)$ Farhi - Susskind model. Our investigation of the $SU(2)$  minimal
walking Technicolor model shows, that the hypercharge assignment can be chosen
in such a way that the theory possesses an additional $Z_{6}$ symmetry.

We also have considered a possible Unification of Technicolor and Standard
Model interactions. It is shown that there is a strong relation between the
monopole content of the Unified Model and the appearance of the additional
discrete symmetry in the Technicolor theory. Namely, the topologically stable
monopoles with the magnetic flux $2 \pi$ cannot appear if the additional
discrete symmetry is absent. Thus if the appearance of such monopoles is
recognized as necessary, then the imposing of the additional discrete symmetry
on Technicolor is preferred.

Let us remind here that the additional discrete symmetry is rather restrictive.
For the Standard Model the requirement that the fermion parallel transporters
are invariant under $Z_6$ provides the important condition for the choice of
the representations, in which the Standard Model fermions may be arranged. So,
if
 the Technicolor model must necessarily preserve the additional discrete
symmetry, we would have an important restriction on the choice of the
Technicolor gauge group. The minimal walking Technicolor model and the $SU(4)$
Farhi - Susskind model satisfy this condition\footnote{Actually, due to the
well - known problems in ETC model building we also do not exclude that the
Technicolor theory, which gives rise to the Electroweak symmetry breaking may
be supplemented with the mechanism of fermion mass generation different from
that of ETC.}.

This work was partly supported by RFBR grants 05-02-16306, 07-02-00237,
08-02-00661, and 09-02-00338.

\clearpage

\end{document}